\RequirePackage{fix-cm}

\documentclass[pdftex,twocolumn,epjc3]{svjour3}          % twocolumn

\RequirePackage[T1]{fontenc}

\smartqed  % flush right qed marks, e.g. at end of proof

\RequirePackage{graphicx}
\RequirePackage{mathptmx}      % use Times fonts if available on your TeX system
\RequirePackage{flushend}
\RequirePackage[colorlinks,citecolor=blue,urlcolor=blue,linkcolor=blue]{hyperref}

\journalname{Eur. Phys. J. C}

\usepackage{graphicx}
\usepackage{amssymb}
\usepackage{subfig}
\usepackage{amsmath}

\begin{document}

\title{Asymptotically flat vacuum solution in modified theory of Einstein's gravity}

%\subtitle{Do you have a subtitle?\\ If so, write it here}

\author{Surajit Kalita\thanksref{e1,addr}
        \and
        Banibrata Mukhopadhyay\thanksref{e2,addr} %etc.
}

%\thankstext[$\star$]{t1}{Thanks to the title}
\thankstext{e1}{e-mail: surajitk@iisc.ac.in}
\thankstext{e2}{e-mail: bm@iisc.ac.in}

\institute{Department of Physics, Indian Institute of Science, Bangalore 560012, India \label{addr}}

\date{Received: 30th August, 2019 / Accepted: 12th October, 2019}
% The correct dates will be entered by the editor

\maketitle

\begin{abstract}
A number of recent observations have suggested that the Einstein's theory of general relativity may not be the ultimate theory of gravity. The $f(R)$ gravity model with $R$ being the scalar curvature turns out to be one of the best bet to surpass the general relativity which explains a number of phenomena where Einstein's theory of gravity fails. In the $f(R)$ gravity, behaviour of the spacetime is modified as compared to that of given by the Einstein's theory of general relativity. This theory has already been explored for understanding various compact objects such as neutron stars, white dwarfs etc. and also describing evolution of the universe. Although, researchers have already found the vacuum spacetime solutions for the $f(R)$ gravity, yet there is a caveat that the metric does have some diverging terms and hence these solutions are not asymptotically flat. We show that it is possible to have asymptotically flat spherically symmetric vacuum solution for the $f(R)$ gravity, which is different from the Schwarzschild solution. We use this solution for explaining various bound orbits around the black hole and eventually, as an immediate application, in the spherical accretion flow around it.
\end{abstract}

\section{Introduction}\label{Introduction}
As the Newtonian theory of gravity falls short to describe various observational data, Einstein's theory of general relativity (GR) becomes the most powerful theory to replace the former. It is undoubtedly the most effective theory to describe the theory of gravity. It can well explain the properties of various compact objects such as black holes, neutron stars, white dwarfs \cite{compact}. It can also explain the various eras of cosmological history of the universe. This theory has already been well tested through various experiments, which Newtonian theory cannot explain, such as deflection of light rays, gravitational redshift, the perihelion precession of Mercury etc. Recently it has again been confirmed through detection of gravitational wave generated from the mergers of binary black holes and neutron stars \cite{2016PhRvL.116f1102A}.

Although GR is one of the most efficient and powerful theories, a number of recent observations have suggested that it may fall short in very high density regions \cite{0004-637X-517-2-565,1538-3881-116-3-1009,0004-637X-607-2-665}. For example, the observations of peculiar type Ia supernovae (SNeIa) either with extremely high luminosity or with extremely low luminosity, which were inferred to be respectively originating from white dwarfs of super-Chandrasekhar limiting mass as high as $2.8M_\odot$ \cite{2006Natur.443..308H,2010ApJ...713.1073S} or from white dwarfs of sub-Chandrasekhar limiting mass as low as $0.5M_\odot$ \cite{1992AJ....104.1543F,1997MNRAS.284..151M,1998AJ....116.2431T,2001PASP..113..308M,2004ApJ...613.1120G,2008MNRAS.385...75T}. In both the scenarios, there is a clear indication of violation of the Chandrasekhar mass-limit. Chandrasekhar mass-limit is the maximum possible mass of white dwarfs (currently accepted value $\sim 1.4M_\odot$ for non-rotating, non-magnetized, carbon-oxygen white dwarfs \cite{1931ApJ....74...81C}) above which the balance due to the force of outward degenerate electron gas and that of inward gravitational pull, no longer sustains, resulting in producing SNeIa. Similarly, a number of neutron stars observed with mass much larger than $2M_\odot$ \cite{2018ApJ...859...54L,0004-637X-728-2-95} are argued to be induced by modified Einstein's gravity \cite{2011JCAP...07..020A,2013JCAP...12..040A}. Moreover, GR cannot explain the era when the size of the universe was smaller than Planck's length. All these observations/inferences suggest that GR may not be the ultimate theory of gravity. Starobinsky was the first who overcame some of these shortcomings in cosmology by means of the modified theory of general relativity \cite{1979ZhPmR..30..719S}. He used the $f(R)$ gravity model, with $R$ being the scalar curvature, to explore some important problems in cosmology. Eventually, a plenty of different models have been proposed to explain various other aspects of observations in astrophysics \cite{Carvalho2017}. Capozziello and his collaborators showed that by means of Starobinsky's $f(R)$ gravity and its higher order corrections, the problem of massive neutron stars can easily be explained \cite{2013JCAP...12..040A,2014PhRvD..89j3509A,2015JCAP...01..001A,2017CQGra..34t5008A,2016IJMPS..4160130A}. Similarly, Mukhopadhyay and his collaborators also showed that these models can also explain both the classes of the white dwarfs, viz. sub- and super-Chandrasekhar limiting mass white dwarfs which produce the peculiar SNeIa \cite{2015JCAP...05..045D,2018JCAP...09..007K}. However none of the above explorations was for the vacuum solution.

The vacuum solution of $f(R)$ gravity is an interesting problem and the 
solutions for a static, spherically symmetric spacetime in $f(R)$ gravity 
were first obtained by Multam\"{a}ki and Vilja \cite{2006PhRvD..74f4022M}. 
They also showed that for a large class models, Schwarzschild-de Sitter metric 
is an exact solution of the field equations. Eventually many researchers have 
obtained a number of solutions for different modified theories of gravity in 
various spacetime geometry. Capozziello and his collaborators obtained 
spherically symmetric vacuum solutions in $f(R)$ gravity using Noether 
symmetry \cite{2007CQGra..24.2153C,2012GReGr..44.1881C}. Later, they also 
obtained axially symmetric vacuum solutions in $f(R)$ gravity considering 
Noether symmetry approach \cite{2010CQGra..27p5008C}. Eventually, similar 
axially symmetric vacuum solutions were also obtained in Weyl's canonical 
coordinates \cite{2013PhLB..718.1493G}. Similarly, spherically symmetric 
solutions of $f(R)$ gravity in the presence of matter were obtained by Shojai 
and Shojai \cite{2012GReGr..44..211S} and these solutions describe the 
equilibrium configuration of a star. Moreover, Vernieri et al. obtained 
anisotropic interior solutions in the presence of Ho\v rava gravity 
\cite{2018EL....12130002V,2018PhRvD..98b4051V}. These new solutions alter 
the event horizon and various important orbits, such as marginally stable, 
marginally bound, photon orbits, etc., and thereby, they change the dynamics 
of the particles moving around the black hole. These solutions have later been used by the researchers to solve various problems of accretion discs \cite{2008PhRvD..78b4043P,2013A&A...551A...4P}. Nevertheless, the solutions, given in these literature, have some diverging terms in the metric components and hence they never reduce to the Schwarzschild metric and thereby to the Minkowski metric at the asymptotic flat limit. This asymptotic flatness is however extremely important in the context of physical problem, e.g. the accretion disc, as a disc extends to a very large region around the compact object and, at the larger radius, no physics should be violated as given by the Schwarzschild or Minkowski metric. In other words, they should pass the solar system tests. Moreover, many of these models assume constant scalar curvature, $R=R_0$ throughout, which is again questionable as for the Schwarzschild metric, $R=0$, and this needs to be satisfied at the asymptotic flat limit. In this paper, we show that the solution for $f(R)$ gravity in vacuum, and hence for black holes, can be obtained which behaves as the Schwarzschild/ Minkowski metric at asymptotic limit and hence this solution can be used in accretion physics effectively.

The paper is organized as follows. In section \ref{f(R) gravity}, we briefly discuss the basic equations of the $f(R)$ gravity, and following in section \ref{solution procedure}, we discuss the possible vacuum solution of these equations. In section \ref{results}, we discuss the behaviour of spacetime obtained for this vacuum solution. We also illustrate various marginal orbits such as marginally stable, marginally bound, photon orbits etc., in case of the $f(R)$ gravity and, eventually, in section \ref{bondi}, we use this solution to explain the spherical accretion flow. At last, we end with conclusions in section \ref{conclusion}.

%=============================================================================

\section{Basic equations in $f(R)$ gravity}\label{f(R) gravity}

Einstein-Hilbert action provides the field equation in general relativity. With the metric signature $(+,-,-,-)$ in 4 dimensions, it is given by \cite{MTW}
\begin{align}\label{Einstein Hilbert Action}
S=\int\Big[\frac{c^4}{16 \pi G}R+\mathcal{L}_M\Big]\sqrt{-g}d^4x,
\end{align}
where $c$ is the speed of light, $G$ the Newton's gravitational constant, $\mathcal{L}_M$ the Lagrangian of the matter field and $g=\det(g_{\mu\nu})$ is the determinant of the metric $g_{\mu\nu}$. Varying this action with respect to $g_{\mu\nu}$ and equating to zero with appropriate boundary conditions, we obtain the Einstein's field equation for general relativity, which is given by
\begin{align}\label{Einstein equation}
G_{\mu\nu} = R_{\mu\nu}- \frac{R}{2}g_{\mu\nu} = \frac{8\pi G}{c^4} T_{\mu\nu} ,
\end{align}
where $T_{\mu\nu}$ is the energy-momentum tensor of the matter field.

In the case of $f(R)$ gravity, the Ricci scalar $R$ is replaced by $f(R)$ in the Einstein-Hilbert action of equation \eqref{Einstein Hilbert Action} resulting in the modified Einstein-Hilbert action, which is given by \cite{2010LRR....13....3D,2017PhR...692....1N}
\begin{align}
S=\int\Big[\frac{c^4}{16 \pi G}f(R)+\mathcal{L}_M\Big]\sqrt{-g}d^4x.
\end{align}
Now varying this action with respective to $g_{\mu\nu}$, with appropriate boundary conditions, we have the modified Einstein equation, which is given by
\begin{equation}\label{modified equation}
\begin{split}
F(R)G_{\mu \nu}+\frac{1}{2}g_{\mu \nu}[RF(R)-f(R)]-(\nabla_\mu \nabla_\nu-g_{\mu \nu}\Box)F(R) \\=\frac{8\pi G}{c^4}T_{\mu \nu},
\end{split}
\end{equation}
where 
\begin{equation}\label{FR}
F(R)=\frac{df(R)}{dR}, 
\end{equation}
$\Box$ is the d'Alembertian operator given by $\Box=\nabla^\mu\nabla_\mu$ and $\nabla_\mu$ is the covariant derivative. For $f(R)=R$, equation \eqref{modified equation} reduces to the Einstein field equation given in equation \eqref{Einstein equation}. For vacuum solution, $T_{\mu \nu} = 0$, which reduces equation \eqref{modified equation} to
\begin{equation}\label{modified vacuum equation}
F(R)G_{\mu \nu}+\frac{1}{2}g_{\mu \nu}[RF(R)-f(R)]-(\nabla_\mu \nabla_\nu-g_{\mu \nu}\Box)F(R)=0.
\end{equation}
The trace of this equation is given by
\begin{equation}\label{modified trace equation}
RF(R)-2f(R)+3\Box F(R)=0.
\end{equation}
By substituting $f(R)$ from this equation in the equation \eqref{modified vacuum equation}, we have
\begin{equation}\label{modified vacuum equation2}
FR_{\mu \nu}-\nabla_\mu \nabla_\nu F=\frac{1}{4}g_{\mu \nu}(RF-\Box F).
\end{equation}
This is the equation which is further used for obtaining the solution in section \ref{solution procedure}.

%=============================================================================
\section{Solution for vacuum spacetime}\label{solution procedure}

We are interested in a spherically symmetric, time independent vacuum solution for the above-mentioned modified Einstein's equation. To obtain that let us choose the spherically symmetric metric $g_{\mu\nu} = \text{diag}(s(r), -p(r), -r^2, -r^2\sin^2\theta)$, where $s,p$ are the functions of the radial co-ordinate $r$ alone. Substituting this metric in the equation \eqref{modified vacuum equation2} and performing some manipulation, we have \cite{2006PhRvD..74f4022M}
\begin{align}\label{equation1}
2\frac{X'}{X}+r\frac{F'}{F}\frac{X'}{X}-2r\frac{F''}{F}&=0
\end{align} and
\begin{align}\label{equation2}
-4s+4X-4rs\frac{F'}{F}+2r^2s'\frac{F'}{F}+2rs\frac{X'}{X}-r^2s'\frac{X'}{X}+2r^2s''&=0,
\end{align}
where $X(r)=p(r)s(r)$ and `prime' denotes derivative with respect to $r$. Let us further assume $F(r)=1+B/r$, such that as $r\to\infty$, $F(r)\to1$, which is the case for GR. Substituting this in the equation \eqref{equation1} and solving for $X(r)$, we obtain
\begin{equation}
X(r)=\frac{C_0r^4}{(B+2r)^4},
\end{equation}
where $C_0$ is the integration constant. As argued before, the solution has to be asymptotically flat, i.e. as $r\to\infty$, $s(r)\to1$ and $p(r)\to 1$. Hence $X(r)\to1$ as $r\to\infty$, which implies that $C_0=16$. Therefore
\begin{equation}
X(r)=\frac{16r^4}{(B+2r)^4}.
\end{equation}
It is evident that for $B=0$, $F(r)=1$ and $X(r)=1$. Substituting $X(r)$ and $F(r)$ in the equation \eqref{equation2} and solving for $s(r)$ along-with expanding it in the power series of $r$, for $B\neq0$, we have
\begin{equation}\label{gtt1}
\begin{split}
s(r)&=\frac{-16+2 B C_1+32 \log2 + (BC_2+8)i\pi}{2 B^2}r^2+1\\&+\frac{B (-24+B C_2)}{24 r}+\frac{B^2-\frac{1}{16} B^3 C_2}{r^2}+\frac{-B^3+\frac{11}{160}B^4 C_2}{r^3}\\&+\frac{188B^4-13 B^5 C_2}{192 r^4}+\dots,
\end{split}
\end{equation}
where $C_1$ and $C_2$ are the integration constants obtained by solving the second order differential equation \eqref{equation2}. As the metric needs to behave as the Schwarzschild metric at large distance, we require the coefficient of $r^2$ to be zero and the coefficient of $1/r$ to be $-2$, which gives
\begin{align*}
C_2 &= \frac{24(B-2)}{B^2}\\
\text{and}\qquad
C_1 &= -8\frac{B(-1+\log4)+(-3+2B)i\pi}{B^2}.
\end{align*}
Therefore, from the equation \eqref{gtt1}, the temporal component of the metric is given by
\begin{equation}\label{gtt}
\begin{split}
g_{tt}=s(r)&=1-\frac{2}{r}-\frac{B(-6+B)}{2 r^2}+\frac{B^2 (-66+13 B)}{20 r^3}\\&-\frac{B^3 (-156+31 B)}{48 r^4}+\frac{3 B^4 (-57+11 B)}{56 r^5}\\&-\frac{B^5 (-360+67B)}{128 r^6}+\dots
\end{split}
\end{equation}
and hence the radial component of the metric is given by $g_{rr} = -p(r) = -X(r)/s(r)$. Moreover, the Ricci scalar or scalar curvature $R$ is given by
\begin{equation}\label{ricci}
\begin{split}
R &= \frac{3 B(-2+B)}{r^4}-\frac{3 B^2(-12+B)}{10 r^5}+\frac{B^3 (-51+8 B)}{10 r^6}\\&-\frac{B^4 (-1776+293 B)}{280 r^7}+\frac{9 B^5 (-944+157B)}{1120 r^8}\\&-\frac{B^6 (-3968+661 B)}{448 r^9}+\dots.
\end{split}
\end{equation}
Since $R$ has to be positive so that gravity has its usual property, i.e. it is always attractive, $B<0$ always. Hence, from equations (\ref{FR}) and (\ref{ricci}), $f(R)$ is given by
\begin{align}
f(R(r)) &= \int F(R) dR \nonumber \\
&= \int F(r) \frac{dR}{dr}dr \nonumber \\
&= \frac{3 B(-2+B)}{r^4}+\frac{3 B^2(-4+7B)}{10 r^5}+\frac{B^3 (-42+ 11B)}{20 r^6}\\&-\frac{B^4 (-552+101 B)}{280 r^7}+\dots \nonumber \\
&= R+K_1R^{5/4}+K_2R^{3/2}+\dots \nonumber\\
&= R+\mathcal{O}(R^{>1}),
\end{align}
with $$K_1 = \frac{12}{5\times3^{5/4}}\frac{B^{3/4}}{(B-2)^{1/4}}, ~~~ K_2=\frac{1}{60\sqrt{3}}\frac{B^{3/2}(B-12)}{(B-2)^{3/2}}.$$
This is the best possible way to represent $f(R(r))$ and it can no longer be written exactly in terms of $R$ only. This is because $R(r)$ is an infinite series of $r$ and hence it cannot be inverted to write $r$ in terms of $R$. In Starobinsky model, the $f(R)$ in Einstein-Hilbert action is considered to be $R^{1+1}$, whereas here it is $R^{1+1/4}$ and higher power of $R$. It is different from the The above form implies that the present gravity is the higher order correction to GR which has many astrophysical and cosmological implications, will be discussed in the next sections.

In case of $B=0$, since $F(r)=X(r)=1$, solving the equations \eqref{equation1} and \eqref{equation2}, we have the Schwarzschild solution, given by
\begin{align}
s(r) = 1-\frac{2}{r}, ~~~ p(r) = \frac{1}{1-\frac{2}{r}}
\end{align}
along-with $R=0$. The solutions for the temporal and radial components, given by equation \eqref{gtt}, show a clear indication of the violation of the Birkhoff's theorem which says that any vacuum solution is essentially the Schwarzschild solution. Hence we can conclude that the Birkhoff's theorem is valid only in the GR spacetime and not in the $f(R)$ gravity regime, which was also discussed earlier for various $f(R)$ gravity models \cite{1979PhRvD..19.2264R,2017AN....338.1015K}.

%=============================================================================
\section{Various properties of the vacuum spacetime for $f(R)$ gravity}\label{results}

In this section, we discuss various physics lying with the vacuum solutions of modified Einstein equation for $f(R)$ gravity. We show that the property of spacetime is same as for the case of the Schwarzschild metric at a large distance.

\subsection{Temporal and spatial components of the metric}
\begin{figure}[!htbp]
     \subfloat[Temporal component, $g_{tt}$\label{temporal}]{
     \centering
       \includegraphics[scale = 0.5]{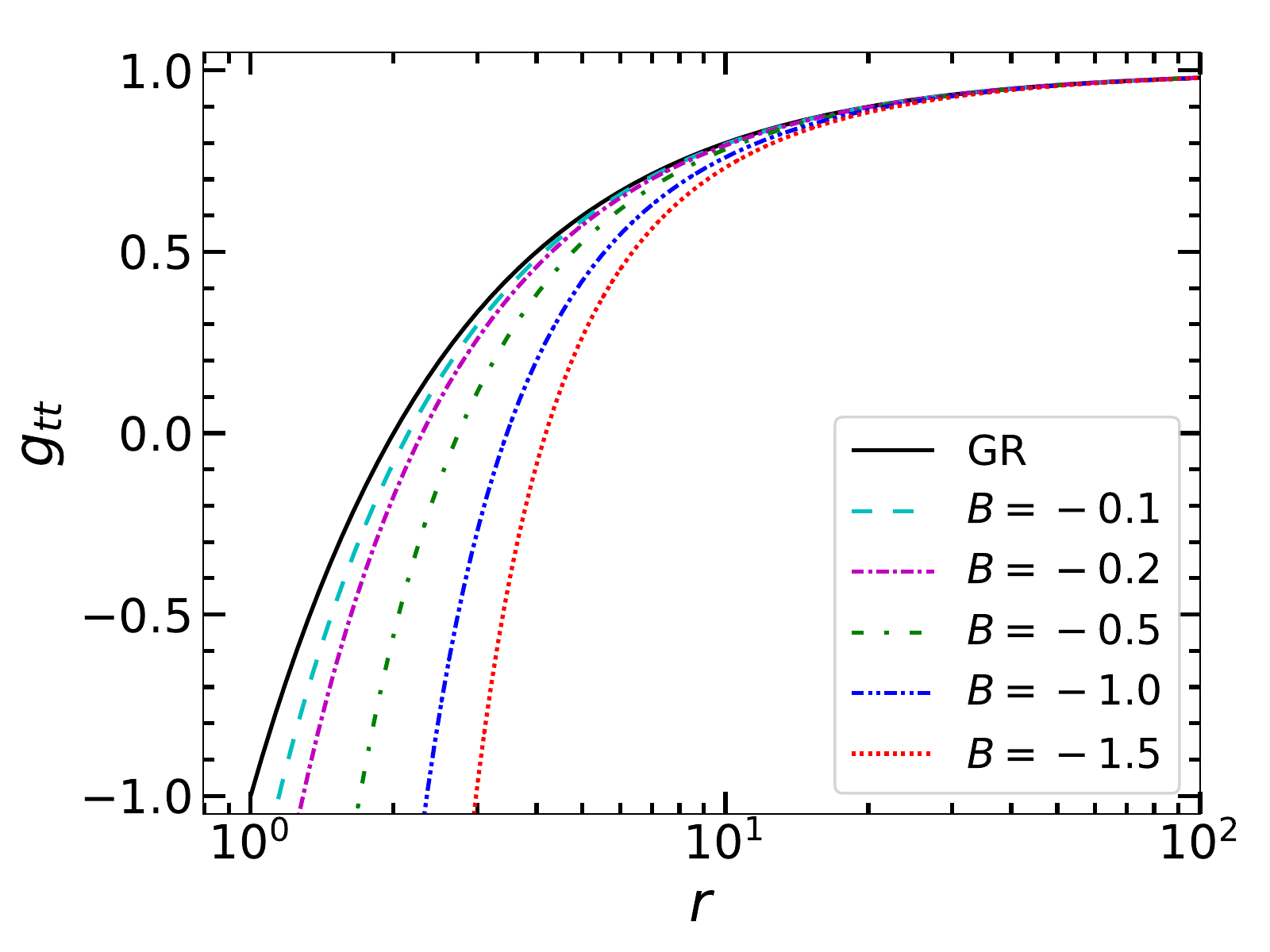}
     }\\
     \hfill
     \subfloat[Radial component, $g_{rr}$\label{radial}]{
     \centering
       \includegraphics[scale = 0.5]{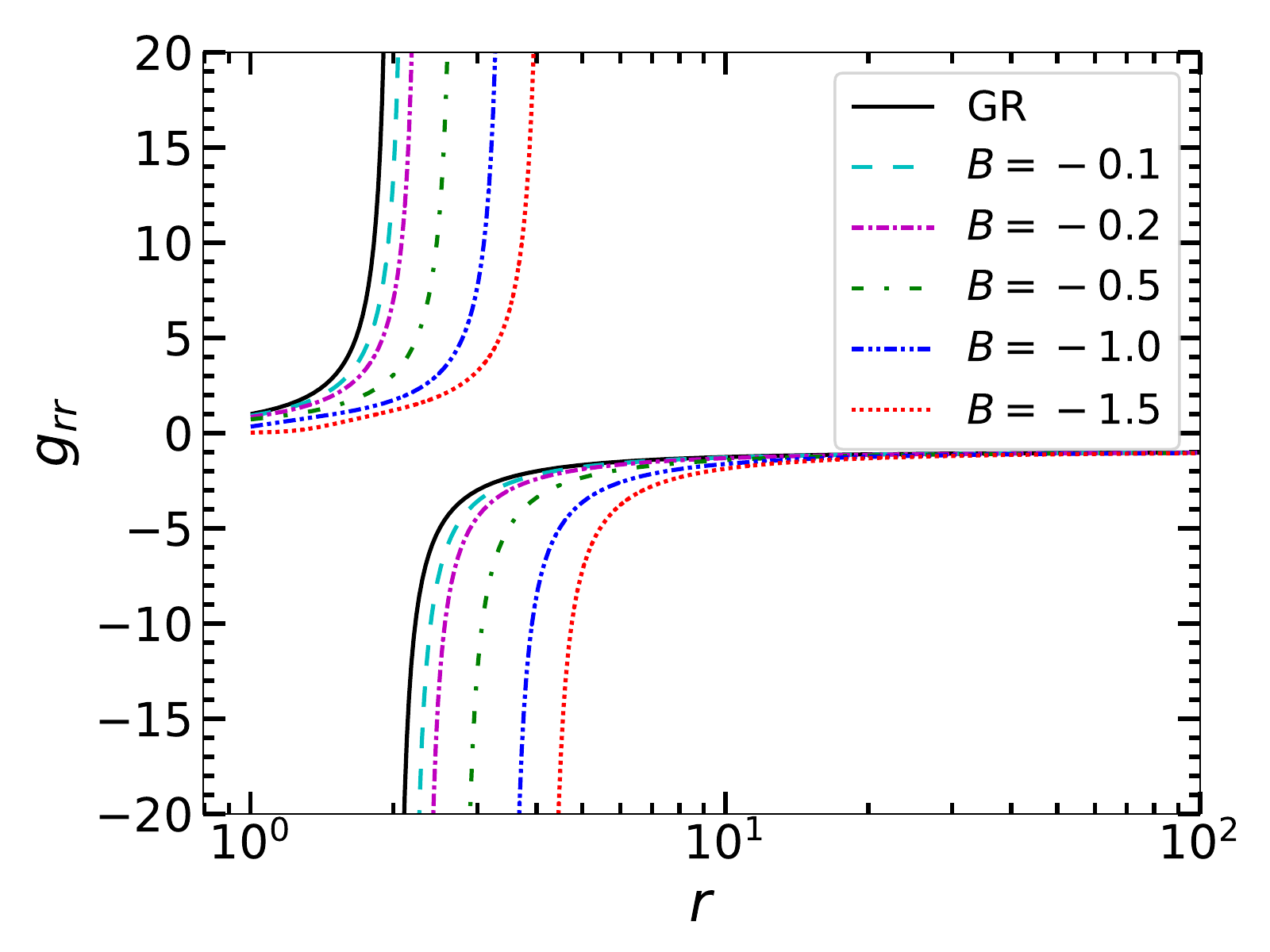}
     }
     \caption{The variation of temporal and radial metric elements as functions of distance $r$. All the quantities are expressed in dimensionless units, which is considered with $c=G=M=1$.}
     \label{temporal and radial}
   \end{figure}

\begin{figure}[!htbp]
    \centering
     \includegraphics[scale = 0.5]{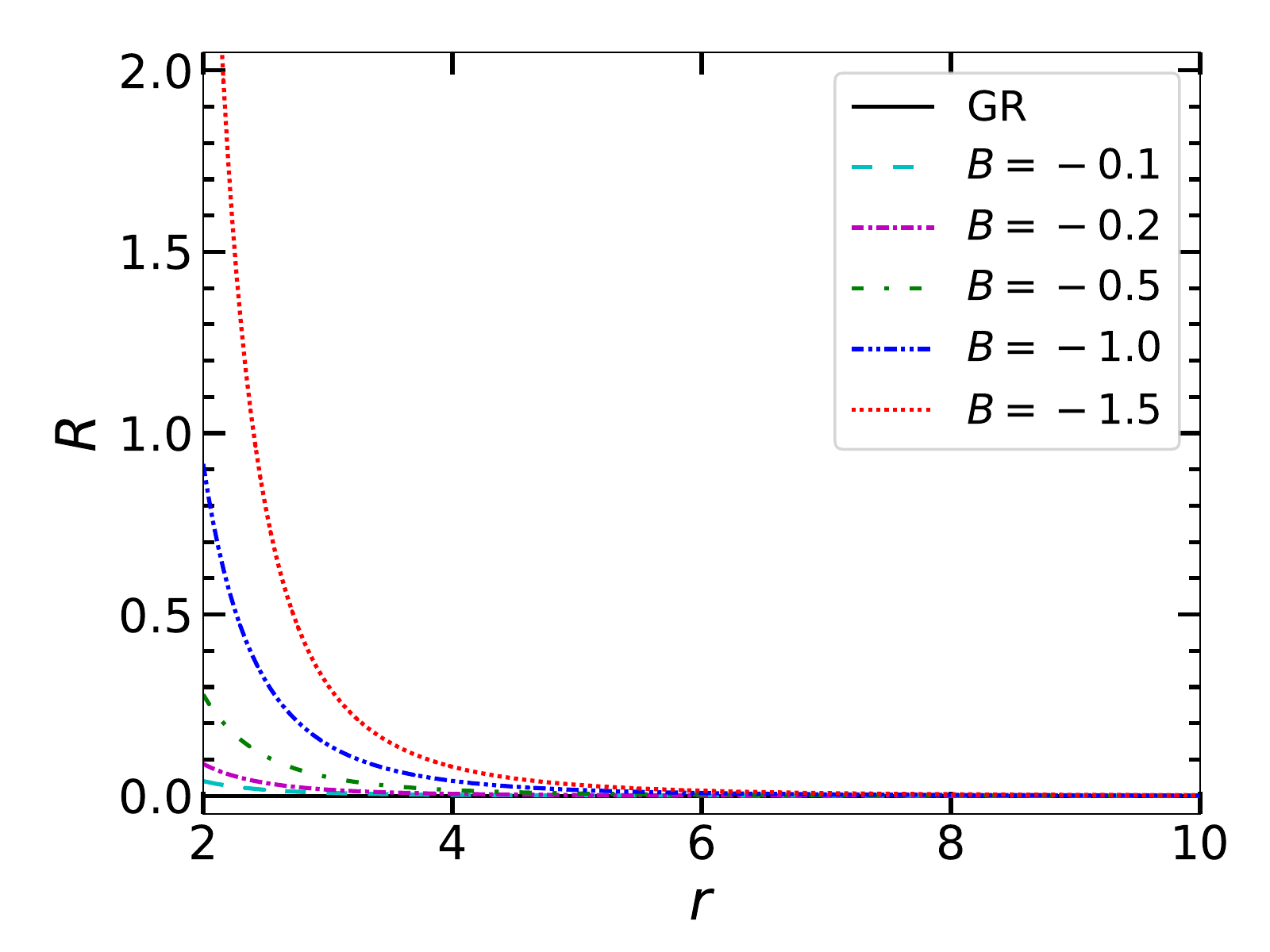}
     \caption{The variation scalar curvature $R$ as a function of $r$. All the quantities are expressed in dimensionless units, which is considered with $c=G=M=1$.}
     \label{Ricci_plot}
   \end{figure}

Figure \ref{temporal and radial} shows the variations of temporal and radial components of the metric as functions of distance $r$ for various values of $B$. Note that $r$ is in the units of $GM/c^2$, where $M$ is the mass of the black hole. From the figure, it is evident that at a large distance, all the curves merge, which implies that all of them tend to the Schwarzschild metric at a large distance. However, near the black hole, there is a significant deviation from the Schwarzschild metric, which reflects the impact of the $f(R)$ gravity therein and its significant effect on the radius of black hole event horizon $r_H$. 
It is also confirmed from Figure \ref{Ricci_plot}, which shows the variation of $R$ with respect to the distance $r$, that at a large distance, $R$ approaches to zero, indicating the Schwarzschild spacetime.

\begin{figure}[!htbp]
    \centering
     \includegraphics[scale = 0.5]{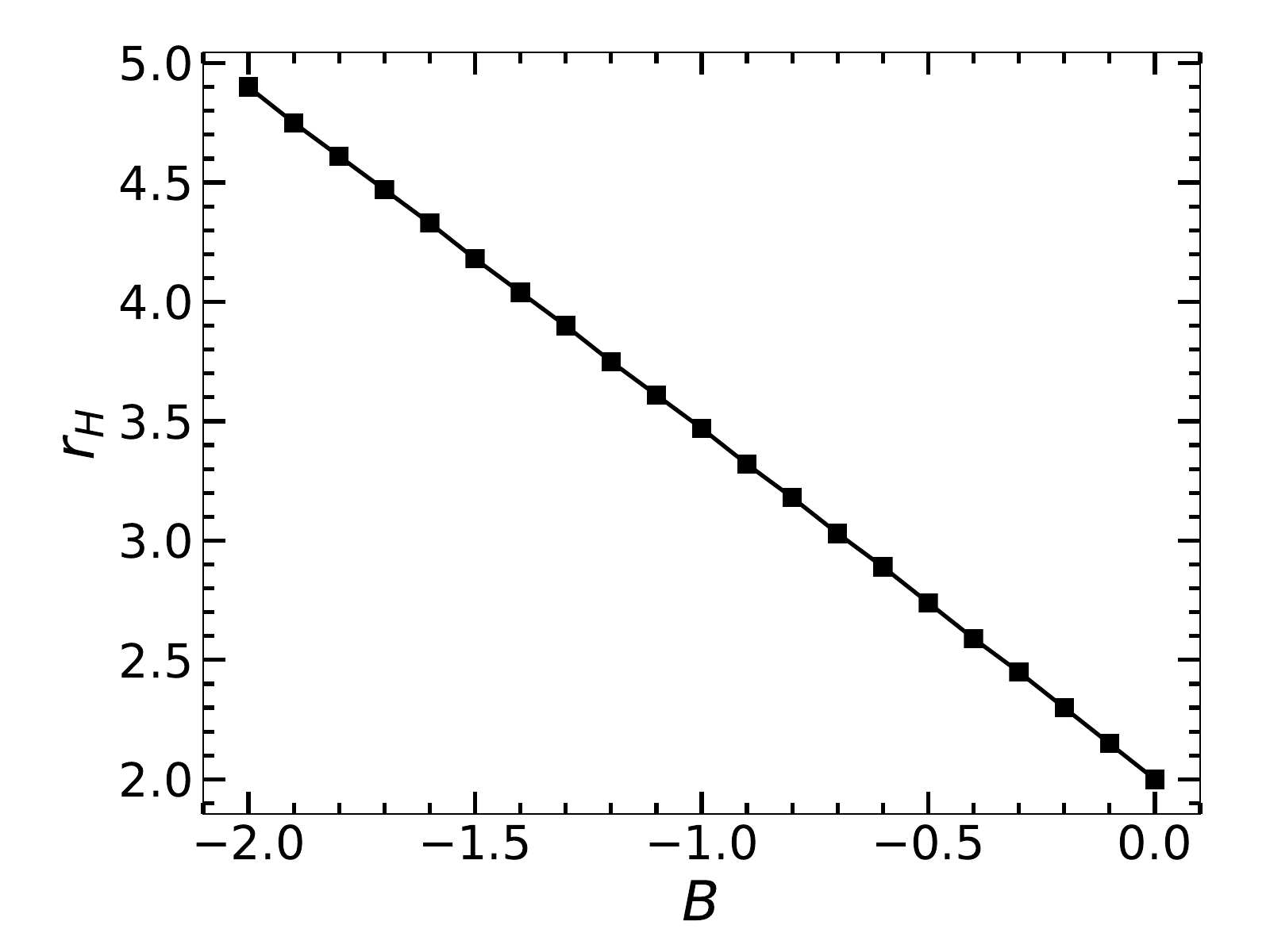}
     \caption{The variation of event horizon $r_H$ as a function of $B$. All the quantities are expressed in dimensionless units, which is considered with $c=G=M=1$.}
     \label{event_horizon}
   \end{figure}

Interestingly, from the divergent nature of $g_{rr}$ (and consequently $g_{tt}$'s 
approaching zero) at smaller radial coordinate $r$ in Figure \ref{temporal and radial}, 
it is evident that with increasing $B$ in magnitude $r_H$ 
increases. It is also depicted in Figure \ref{event_horizon}. It confirms the impact of
$f(R)$ gravity on the size of black hole for the same mass as the Schwarzschild case.
In GR, the size of $r_H$ is completely determined by $M$ for a non-rotating black hole.
However, above fact implies that in the $f(R)$ gravity premise, even a non-rotating
black hole radius is determined by additional metric parameter(s), depending on the property of $f(R)$.

%--------------------------------------------------------------------------------

\subsection{Marginally stable and bound orbits in $f(R)$ gravity}

\begin{figure}[!htbp]
     \subfloat[marginally bound\label{bound}]{%
     \centering
       \includegraphics[scale = 0.5]{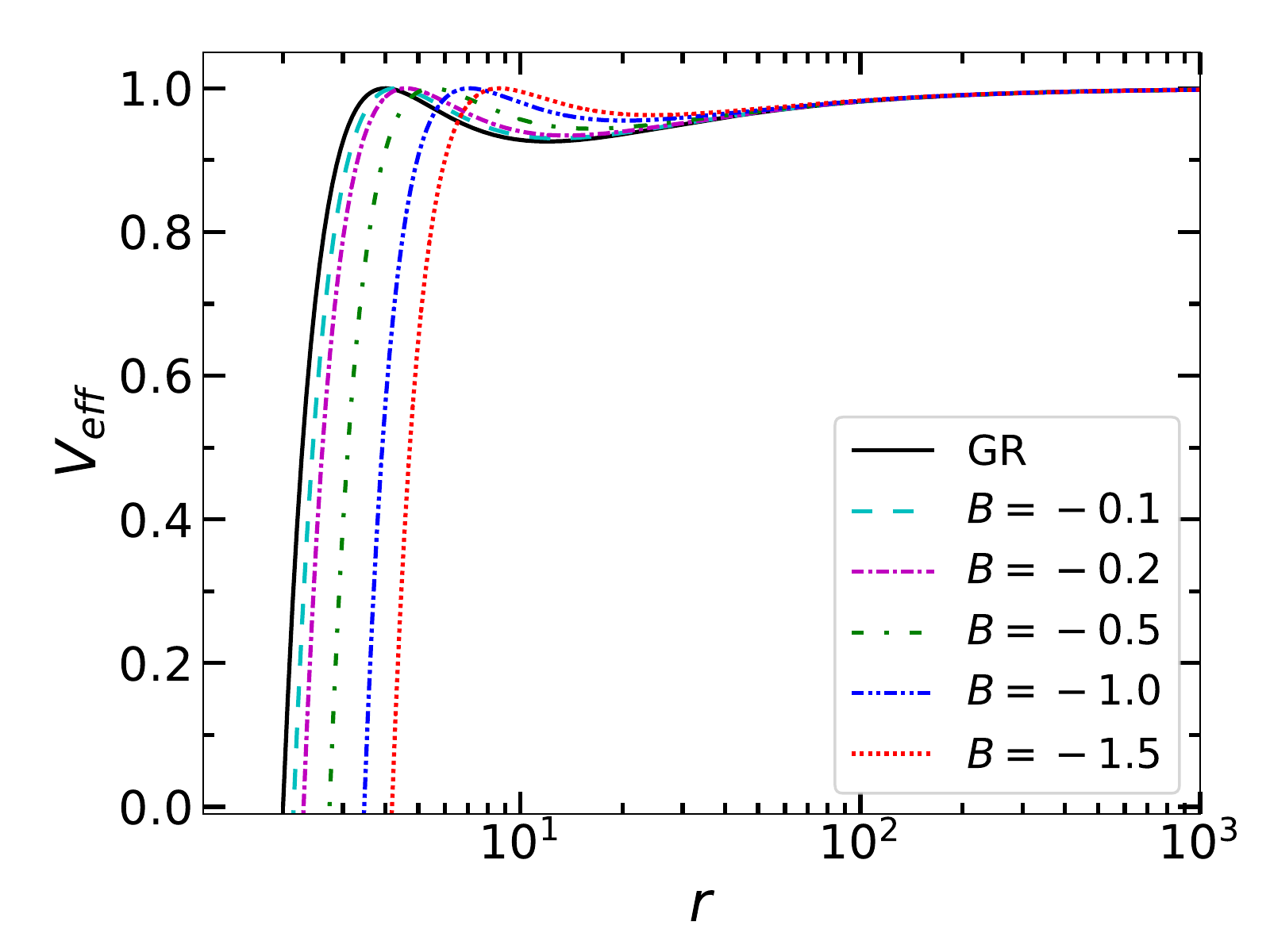}
     }
     \hfill
     \subfloat[marginally stable\label{stable}]{%
     \centering
       \includegraphics[scale = 0.5]{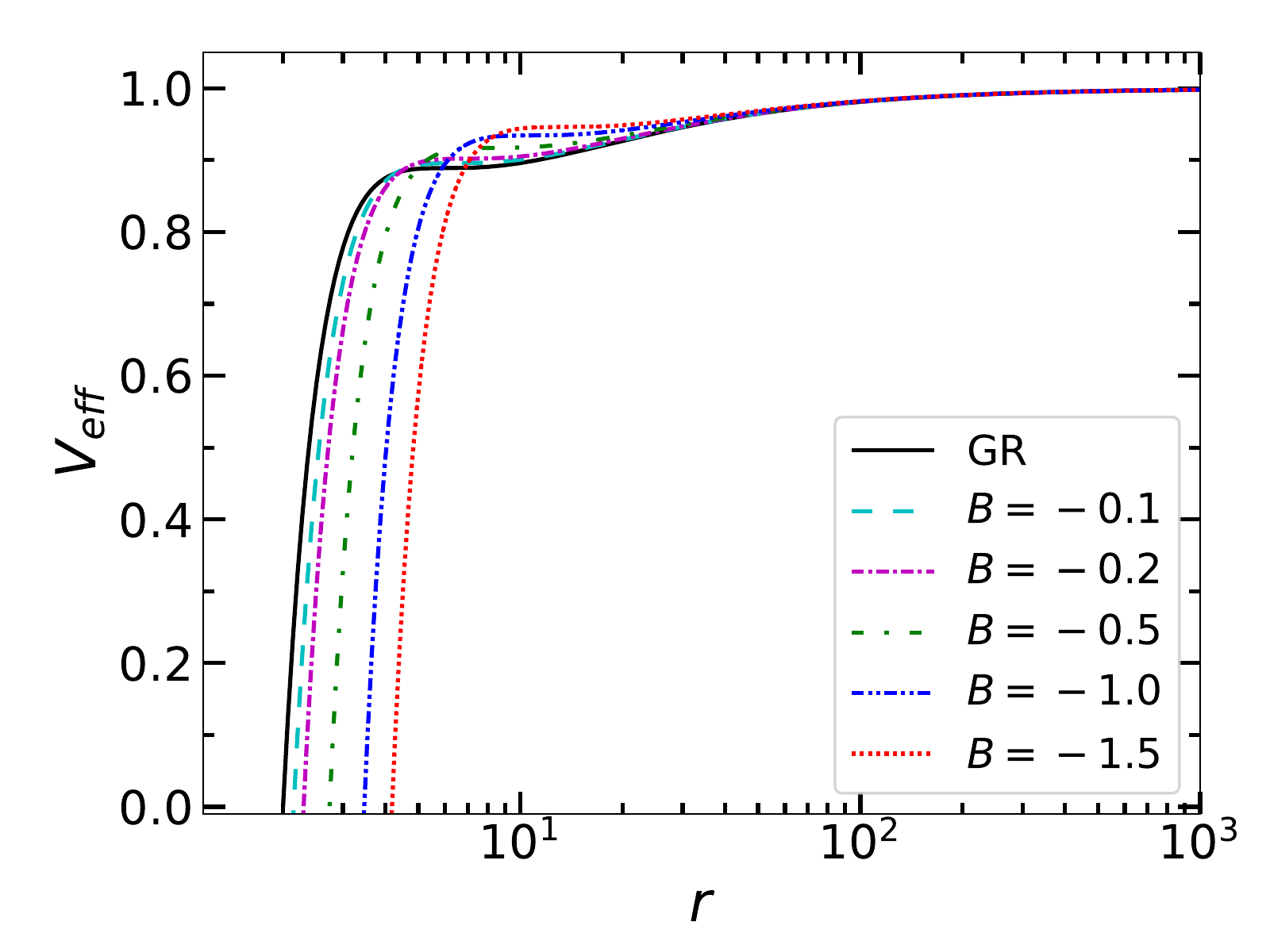}
     }
     \caption{Effective potential for marginally bound and marginally stable circular orbits. All the quantities are expressed in dimensionless unit with $c=G=M=1$.}
     \label{marginal orbits}
   \end{figure}

Here we explore various orbits of a test particle motion around the black hole. The conditions required for the marginally stable circular orbit, marginally bound circular orbit and photon orbit for a spherically symmetric metric of the form $g_{\mu\nu} = \text{diag}(e^{2\phi(r)}, -e^{2\lambda(r)},-r^2, -r^2\sin^2\theta)$ are respectively given by
\begin{align}
1-e^{2\phi(r)}-r\frac{d\phi}{dr}&=0,\\
3\frac{d\phi}{dr}-2r\Big(\frac{d\phi}{dr}\Big)^2+r\frac{d^2\phi}{dr^2}&=0,\\
r\frac{d\phi}{dr}-1&=0.
\end{align}
Note that the common condition to obtain these equations is the minimization of the effective potential. The other conditions are the marginal stability for the marginally stable circular orbit, marginal boundness for marginally bound circular orbit and maximization of the effective potential for the photon orbit. On the other hand, the effective potential for a massive particle is given by $V_{eff} = g_{tt}(1+L^2/r^2)$, whereas for massless particle like photon, it is given by $V_{eff} = g_{tt} L^2/r^2$, with $L$ being the specific angular momentum of the particle. For a massive particle, $L$ is given by
\begin{align}
L = \sqrt{\frac{r^3 \phi'(r)}{1-r\phi'(r)}},
\end{align}
and hence the total specific energy is given by
\begin{align}
E = \sqrt{1+\frac{L^2}{r^2}}e^{\phi(r)} = \frac{e^{\phi(r)}}{\sqrt{1-r\phi'(r)}}.
\end{align}
Here for convenience, we assume $c=G=M=1$. Table \ref{parameters of space-time} shows various marginal orbits for different values of $B$ and Figure \ref{marginal orbits} shows $V_{eff}$ for marginally bound and marginally stable circular orbits for various values of $B$. Here GR represents nothing but the results in the Schwarzschild spacetime. It is interesting to note that as $B$ increases, $r_{H}$ increases and, as a result, the radii of all the marginal orbits increase.

\begin{table}[!htp]
\centering
\caption{Various parameters of spacetime for different values of $B$: $r_{H}$ is the event horizon, $r_{MB}$ the marginally bound orbit, $r_{MS}$ the marginally stable orbit, $L_{MB}$ and $L_{MS}$ are their corresponding specific angular momenta and $r_{ph}$ is the photon orbit. All the values are in dimensionless unit considering $c=G=M=1$.}
\label{parameters of space-time}
\begin{tabular}{|l|l|l|l|l|l|l|}
\hline
$B$ & $r_{H}$ & $r_{MB}$ & $r_{MS}$ & $r_{ph}$ & $L_{MB}$ & $L_{MS}$\\
\hline\hline
GR & 2.00 & 4.00 & 6.00 & 3.00 & 4.00 & 3.46\\ 
-0.1 & 2.15 & 4.30 & 6.45 & 3.20 & 4.15 & 3.61\\ 
-0.2 & 2.30 & 4.60 & 6.90 & 3.40 & 4.29 & 3.75\\ 
-0.5 & 2.74 & 5.52 & 8.28 & 3.98 & 4.71 & 4.15\\ 
-1.0 & 3.47 & 7.07 & 10.64 & 4.94 & 5.37 & 4.78\\ 
-1.5 & 4.18 & 8.66 & 13.08 & 5.89 & 5.99 & 5.37\\ 
\hline
\end{tabular}
\end{table}

%=================================================================================
\section{Spherical accretion flow in $f(R)$ gravity}\label{bondi}

In this section, we explore the effect of above spacetime solution in the spherical accretion flow. Bondi introduced spherical accretion in the Newtonian framework in which matter flows radially to the central object without having any angular momentum \cite{1952MNRAS.112..195B}. Eventually the spherical accretion problem was solved in the Schwarzschild spacetime \cite{1972Ap&SS..15..153M}. We use here similar technique to investigate the effect of $f(R)$ gravity in the spherical accretion flow.

Let us consider the static spherically symmetric spacetime metric as $g_{\mu\nu} = \text{diag}(-e^{2\phi(r)},e^{2\lambda(r)},r^2,r^2\sin^2\theta)$. The velocity gradient equation of the flow is given by (equivalent equations for the Schwarzschild geometry are given in \cite{1972Ap&SS..15..153M})
\begin{equation}
\begin{split}
\frac{du}{u}\Bigg[V^2-\frac{u^2}{u^2+e^{-2\lambda}}\Bigg]+\frac{dr}{r}\Bigg[2V^2+r(V^2-1)(\phi'+\lambda')\\+\frac{r \lambda'e^{-2\lambda}}{u^2+e^{-2\lambda}}\Bigg] = 0,
\end{split}
\end{equation}
where $u = dr/dt$, $V^2 = 4T/3(1+4T)$ with $T$ being the temperature of the fluid which is defined as $T \equiv P/\rho$, where $P$ and $\rho$ are respectively the pressure and density of the fluid. The adiabatic equation of state is considered here, which is given by $P\propto\rho^\gamma$ with $\gamma$ being the adiabatic index. Assuming the fluid mostly contains hot relativistic ions, we choose $\gamma=4/3$.

   \begin{figure}[!htbp]
     \subfloat[$T_{out} = 10^4$ K]{%
     \centering
       \includegraphics[scale = 0.5]{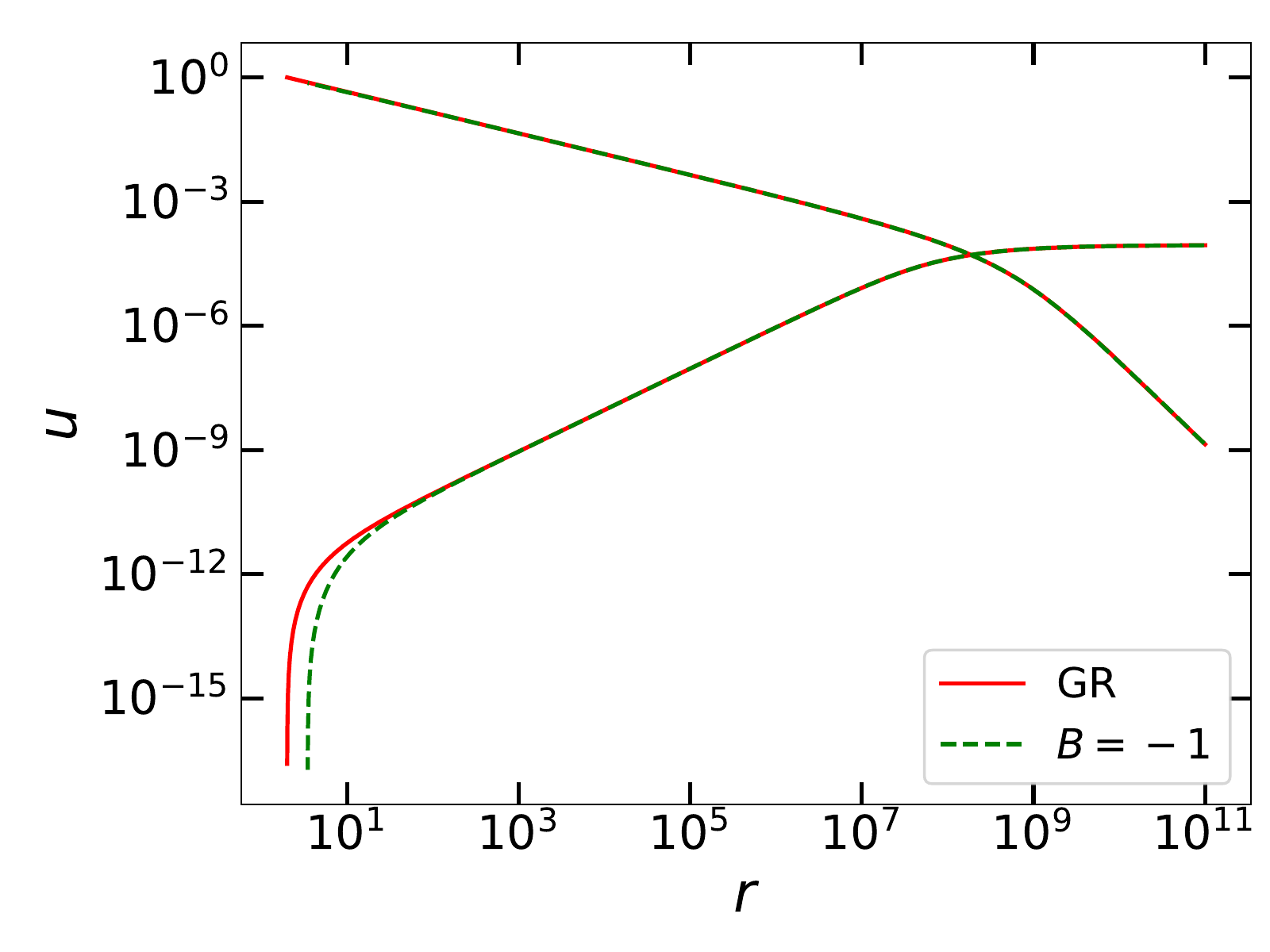}
     }
      \hfill
     \subfloat[$T_{out} = 10^8$ K]{%
     \centering
       \includegraphics[scale = 0.5]{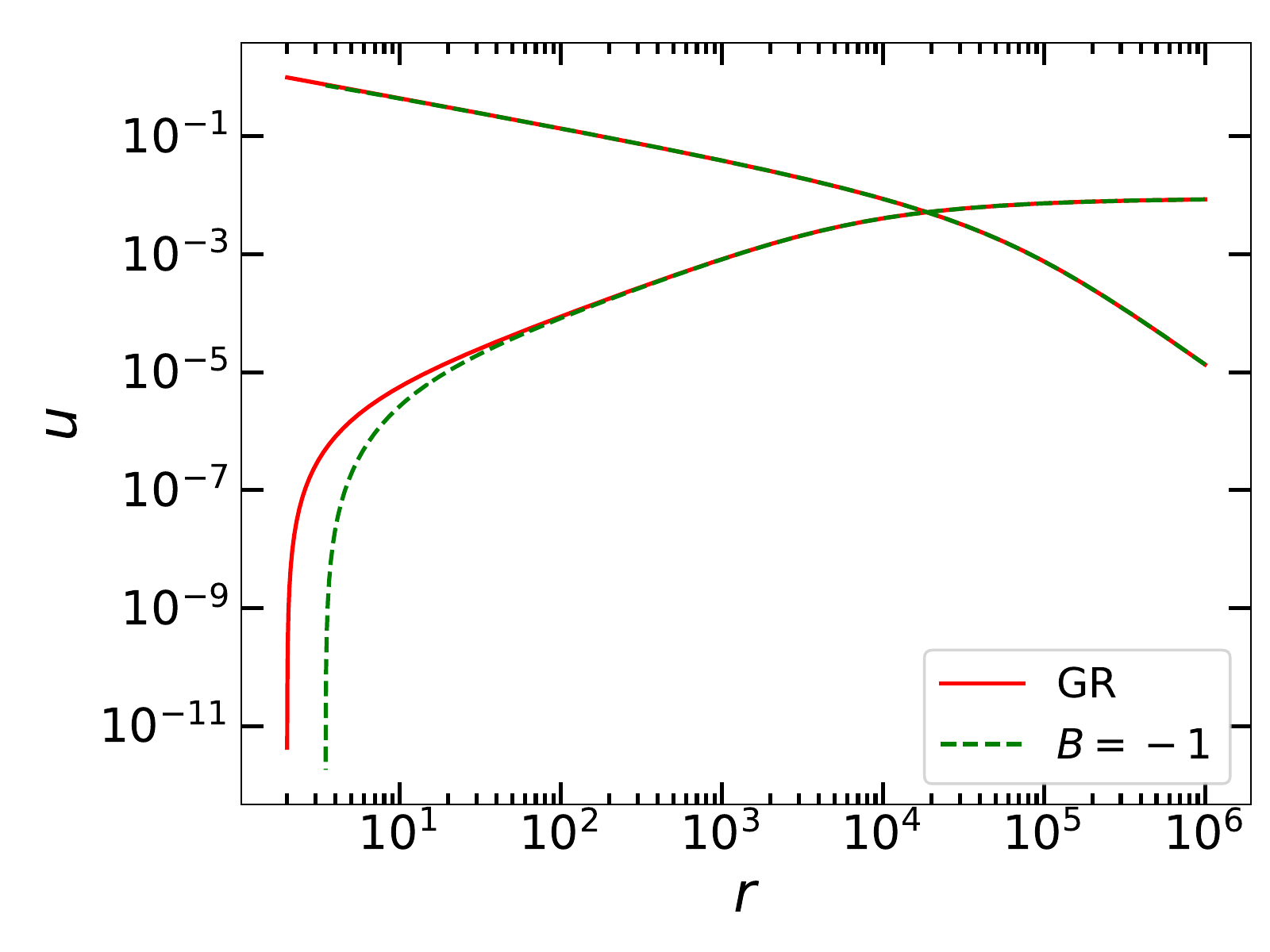}
     }
      \hfill
     \subfloat[$T_{out} = 10^{11}$ K]{%
     \centering
       \includegraphics[scale = 0.5]{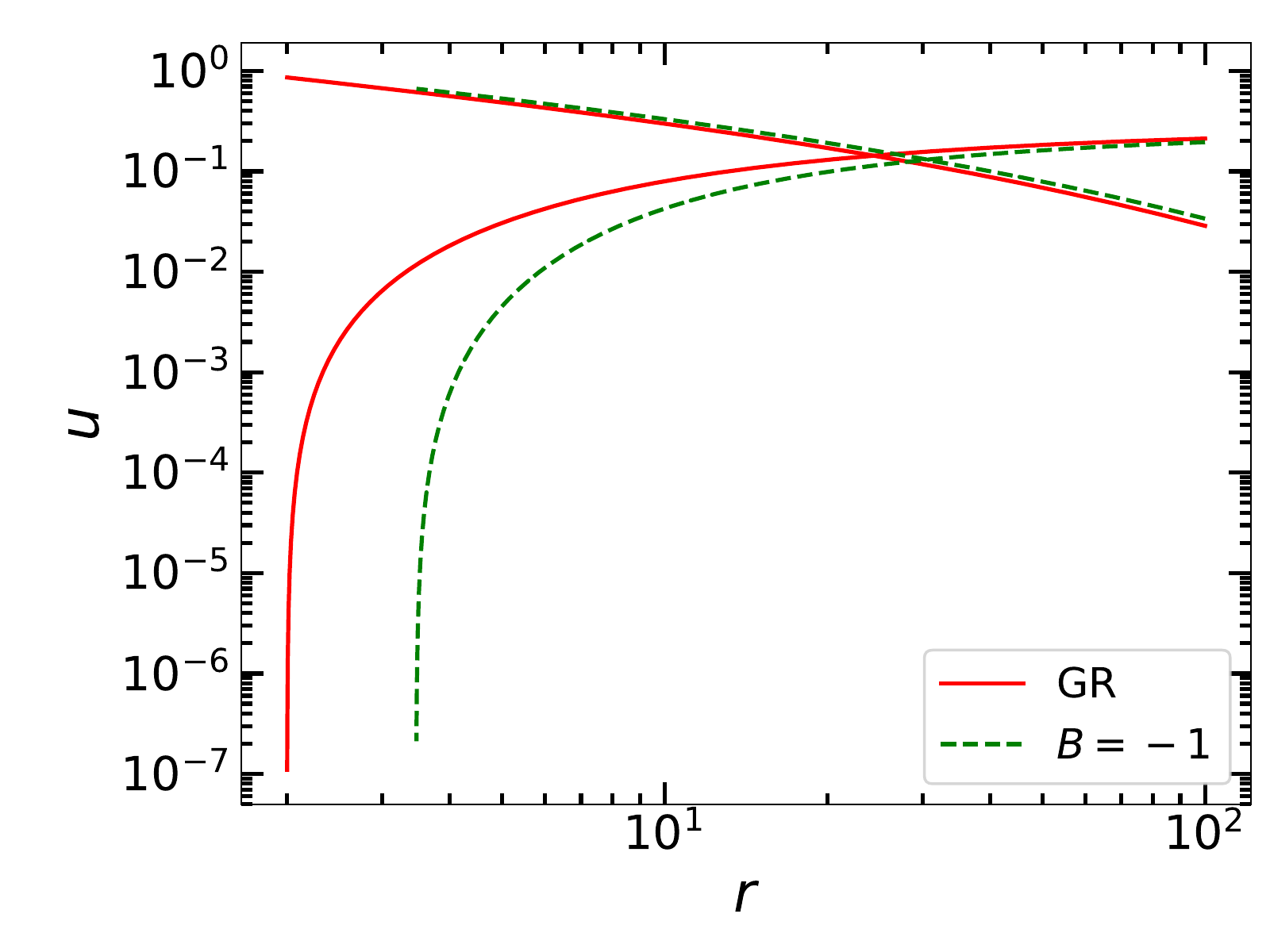}
     }
     \caption{Spherical accretion flow in modified gravity: red solid line corresponds to the Schwarzschild spacetime and green dashed line corresponds to the $f(R)$ gravity with $B=-1$. The left panel is obtained considering outside temperature $10^4$ K, middle panel is for $10^8$ K and the right panel corresponds to $10^{11}$ K. All the quantities are expressed in dimensionless unit with $c=G=M=1$.}
     \label{bondi_plot}
   \end{figure}
Figure \ref{bondi_plot} illustrates the accretion and wind flows for the spherical accretion in $f(R)$ gravity. We assume that the matter starts exhibiting spherical accretion flow, once the Keplerian disc flow ends. In other words, we assume that as the matter comes close enough to the black hole, it loses all its angular momentum, resulting in radial fall to the black hole. Of course, in reality, such flow will be advective accretion flow with non-zero angular momentum. However, here in the first approximation, as an immediate simpler application of our $f(R)$ gravity solution, we assume the flow to be spherical. The three panels, shown in Figure \ref{bondi_plot}, corresponds to three different temperatures ($T_{out}$) at which matter starts behaving like the spherical accretion flow. Note that, the sonic point radius remains the same as for the case of Schwarzschild spacetime and it is located very far from the black hole, if $T_{out}$ is very small. However, since the event horizon shifts in case of the modified gravity, both the accretion and wind branches are deviated from those in the case of Schwarzschild spacetime, close to the central object. At $r = r_{H}$, the velocity of accreting particle reaches the velocity of the light, whereas the wind particle has a very low speed near $r_{H}$ and it starts gaining speed as the radius increases. On the other hand, if $T_{out}$ is large enough, the sonic point corresponding to the Schwarzschild spacetime and that for the $f(R)$ gravity differ significantly. This model is of course a very simplistic model, but we use it just to illustrate the imprint of the modified gravity. It however seems that the $f(R)$ gravity does not have significant practical effects on the spherical accretion flow, which means that Einstein's gravity is sufficient in order to explain the spherical accretion flow. A better exploration in a realistic model containing angular momentum profile, e.g. accretion discs, will be carried out in future.

%=================================================================================

\section{Conclusion}\label{conclusion}
In the literature, it has already been discussed about the behaviour of vacuum spacetime as well as various marginal orbits in the context of $f(R)$ gravity. However, the main caveat in those models is the consideration of constant scalar curvature $R$, due to which the temporal and radial components of the metric turn out to be diverging at a large distance. In other words, the metric is not asymptotically flat. In this paper, we have explicitly shown that we can still obtain asymptotically flat vacuum spacetime metric in the context of $f(R)$ gravity of form $R+\mathcal{O}(R^{>1})$. The particular form of additional term $\mathcal{O}(R^{>1})$ plays the main role in determining properties of spacetime deviated from GR, while $R$ corresponds to the GR effect. Nevertheless, this form of $f(R)$ is similar to those proposed by Starobinsky ($\mathcal{O}(R^{>1})=R^2$) \cite{1979ZhPmR..30..719S} in cosmology to explain acceleration expansion of the universe, ourselves earlier (e.g., $\mathcal{O}(R^{>1})=\alpha R^2(1-\gamma R)$ and $\alpha R^2 e^{-\gamma R}$) \cite{2018JCAP...09..007K} in astrophysics to explain peculiar over- and under-luminous SNeIa, and others (e.g., $\mathcal{O}(R^{>1})=\gamma R^2 + \beta R^3$) \cite{2014PhRvD..89j3509A} in various astrophysical and cosmological contexts. There are many properties associated with black hole sources, e.g. quasi-period oscillation, whose origins remain (completely) unresolved in GR. The presently proposed asymptotically flat $f(R)$ gravity might be very useful to enlighten these issues.

It is achieved on consideration of varying scalar curvature which vanishes in the limit $r\to\infty$, giving rise to the asymptotic flat spacetime metric. Hence, the effect of modified gravity reduces to that of general relativity and eventually of the Minkowski spacetime, far away from the black hole. We have also argued that this is a clear indication of the violation of Birkhoff's theorem in presence of the modified gravity.

To investigate the effect of this spacetime, we have first explored the properties of various marginal orbits. We have shown that the radii of various orbits as well as the event horizon shift in modified gravity premise. This deviation is prominent when the deviation in the $f(R)$ is more compared to that of general relativity. We have further investigated the effect of $f(R)$ gravity in the spherical accretion flow. Here also, we have found that the physics of the spherical accretion remains same as that for the case of general relativity at a very large distance from the black hole. However, since the event horizon shifts in the case of modified gravity, the properties of the accretion as well as wind flows change close to the central object, although the change is not very significant for practical purpose. To summarize, we argue that it is possible to obtain physically viable vacuum solution in the case of $f(R)$ gravity, which can be used for further applications in astrophysics.

\bibliographystyle{spphys}
\bibliography{mypaper3}

\end{document}